\documentclass[final,5p,times,twocolumn]{elsarticle}

\journal{Physics Letters A}

\begin{document}
\begin{frontmatter}

\title{Isotope shift of the ferromagnetic transition temperature in itinerant
ferromagnets
}

\author{Takashi Yanagisawa, Izumi Hase and Kosuke Odagiri
}

\address{Electronics and Photonics Research Institute,
National Institute of Advanced Industrial Science and Technology (AIST),
Tsukuba Central 2, 1-1-1 Umezono, Tsukuba 305-8568, Japan
}

\begin{abstract}
We present a theory of the isotope effect of the Curie temperature $T_{\rm c}$ 
in itinerant ferromagnets.
The isotope effect in ferromagnets occurs via the electron-phonon vertex 
 correction and the effective attractive interaction mediated by the
electron-phonon interaction.
The decrease of the Debye frequency increases the relative strength
of the Coulomb interaction, which results in a positive isotope
shift of $T_{\rm c}$ when the mass $M$ of an atom increases.
Following this picture, we evaluate the isotope effect of $T_{\rm c}$ by using the
Stoner theory and a spin-fluctuation theory.
When $T_{\rm c}$ is large enough as large as or more than 100K, the isotope
effect on $T_{\rm c}$ can be measurable.
Recently, precise measurements on the oxygen isotope effect on $T_{\rm c}$
have been performed for itinerant ferromagnet SrRuO$_3$ with 
$T_{\rm c}\sim 160$K.
A clear isotope effect has been observed with the positive shift of 
$T_{\rm c}\sim 1$K by isotope substitution ($^{16}O\rightarrow ^{18}O$).
This experimental result is consistent with our theory.
\end{abstract}

\begin{keyword}
itinerant ferromagnet; isotope effect; Hubbard model;
electron-phonon interaction; vertex correction;
Stoner theory; spin-fluctuation theory

\PACS{75.10.Lp, 75.47.Lx, 75.50.Cc}
\end{keyword}

\end{frontmatter}

\section{Introduction}

Strongly correlated electron systems (SCES) have been investigated
intensively, because SCES  exhibit many interesting quantum
phenomena. 
SCES include, for example, cuprate high-temperature 
superconductors\cite{bed86,benn,and97,dag94},
heavy fermions\cite{ste84,ott87,map00,kondo}, and organic conductors\cite{ish98}.
In the study of magnetism, the Hubbard model is regarded as one of the most 
fundamental
models\cite{mor85,hub63,hub64,gut63,kan63,yan96,yam98,yan16}.
The electron-phonon interaction is also important in metals and even in correlated 
electron systems.
The electron-phonon interaction has a ubiquitous presence in materials.
%An unconventional isotope effect has been reported for high-temperature
%cuprate superconductors\cite{mul90, zha01}.
%There is also an unconventional isotope effect in iron-based
%superconductors\cite{shi09,yan09}.

The isotope effect of the ferromagnetic transition has been investigated
for several materials.
They are La$_{1-x}$Ca$_x$MnO$_3$\cite{fra98,fra98b}, 
Pr$_{1-x}$Ca$_x$MnO$_3$\cite{fis03}, 
RuSr$_2$GdCu$_2$\cite{pri99}, ZrZn$_2$\cite{kna70} and SrRuO$_3$\cite{kaw16}. 
First three compounds La$_{1-x}$Ca$_x$MnO$_3$, Pr$_{1-x}$Ca$_x$MnO$_3$
and RuSr$_2$GdCu$_2$ show that $T_{\rm c}$ decreases upon the
isotope substitution $^{16}$O$\rightarrow ^{18}$O.
The isotope shift of $T_{\rm c}$ for ZrZn$_2$ was not determined
because the shift of $T_{\rm c}$ is very small and there was uncertainty
arising from different impurity levels.
The compound SrRuO$_3$ exhibits a positive isotope shift, that is,
$T_{\rm c}$ increases upon $^{18}$O isotope substitution.
We think that mechanisms of the isotope effect for the first three
materials and the last one SrRuO$_3$ are different.

The large Curie temperature shift $T_{\rm c}$($^{16}$O) = 222.7K to
$T_{\rm c}$($^{18}$O) = 202.0K was reported when $x=0.20$ for 
La$_{1-x}$Ca$_x$MnO$_3$\cite{fra98,fra98b}. 
We consider that
this shift is caused by strong electron-lattice coupling with some
relation to large magnetoresistance\cite{vol54,jin94}.
There is a suggestion that the ferromagnetic transition is caused by the
double-exchange interaction\cite{zen51,goo55,and55} and a strong
electron-lattice interaction originating from the Jahn-Teller effect\cite{mil95}.
Pr$_{1-x}$Ca$_x$MnO$_3$ is also a member of materials
that exhibit the colossal magnetoresistance phenomenon\cite{fis03}.
The Curie temperature was lowered due to the isotope substitution
$^{16}$O$\rightarrow$$^{18}$O; $T_{\rm c}$($^{16}$O) = 112K is shifted
to $T_{\rm c}$($^{18}$O) = 106K when $x=0.2$.
It is expected that the isotope effect arises from the same 
mechanism as for La$_{1-x}$Ca$_x$MnO$_3$\cite{zha96,zha99}.

As for strontium ruthenates, 
Raman spectra of SrRuO$_3$ films showed anomalous temperature
dependence near the ferromagnetic transition temperature\cite{kir95}.
This indicates that the electron-phonon interaction plays a role in
SrRuO$_3$.
Recently, the isotope effect of the Curie temperature $T_{\rm c}$ has been
reported in SrRuO$_3$\cite{kaw16}.
This material is an itinerant ferromagnet with $T_{\rm c}\simeq 160$K.
The ferromagnetic transition temperature was increased about 1K
upon $^{18}$O isotope substitution.
A softening of the oxygen vibration modes is induced by the isotope
substitution ($^{16}$O$\rightarrow$$^{18}$O).  This was clearly
indicated by Raman spectroscopy. The Raman spectroscopy also confirmed
that almost
all the oxygen atoms (more than 80 percent) were substituted
successfully.
The increase of the atomic mass leads to a decrease of the Debye
frequency $\omega_{\rm D}$.  In fact,
the Raman spectra clearly indicate that the main vibration frequency
of $^{16}$O at 372cm$^{-1}$
is lowered to 351cm$^{-1}$ for $^{18}$O by oxygen isotope substitution
in SrRuO$_3$.
This shift is consistent with the formula
$\omega_{\rm D}\propto 1/\sqrt{M}$ where $M$ is the mass of an oxgen atom.
Thus, experiments confirmed that the isotope shift of $T_{\rm c}$ is induced
by the decrease of the frequency of the oxygen vibration mode.

In this paper we investigate the isotope shift of the Curie temperature
theoretically.
The paper is organized as follows.
In the next Section, we outline the theory of isotope effect in a
ferromagnet.
In the Section 3 we show the Hamiltonian.  In the Section 4, we examine the
corrections to the ferromagnetic state due to the electron-phonon interaction,
by examining the ladder, self-energy and vertex corrections.
In the Section 5, we calculate the oxygen-isotope shift of $T_c$ on the basis of
the spin-fluctuation theory. 
We show that the both theories give consistent results on the isotope effect.

\section{Isotope effect in a ferromagnet}

The reduction of the Debye frequency results in the increase of relative strength
of the Coulomb interaction $U$.  This results in a positive isotope
shift of $T_c$.
This is a picture that indicates the positive isotope shift of
$T_{\rm c}$; $\partial T_c/\partial M>0$.

We start from the Hubbard model with the on-site Coulomb repulsion $U$ to 
describe a ferromagnetic state.
The Curie temperature $T_{\rm c}$ is determined by the gap equation.
The effective attractive interaction due to the phonon exchange reduces $U$ 
to $U+g$ ($g<0$) in the neighborhood of the Fermi surface. 
The effective attraction, however, shows no isotope shift in the Stoner theory 
because the Curie temperature 
is determined by the interaction at the Fermi surface
and then the variation of
$\omega_{\rm D}$ has no effect on $T_{\rm c}$.
The electron-phonon vertex correction reduces the magnetization and this leads
to the isotope effect.  
Although the vertex correction is on order of $\omega_{\rm D}/\epsilon_{\rm F}$, for
the Debye frequency $\omega_{\rm D}$ and the Fermi energy $\epsilon_{\rm F}$, the
isotope effect can be observed by precise measurements when the Curie temperature
is as large as 100K or more than that. 

The isotope effect in itinerant ferromagnets was
first investigated on the basis of the Stoner theory in Ref.\cite{app80}, and
the formula for isotope coefficient $\alpha$ was given.
A fluctuation effect, however, is not included in the Stoner theory.
Because the spin-fluctuation theory has been successful in understanding
physical properties in itinerant ferromagnets\cite{mor85},
a formula based on the spin-fluctuation theory is necessary. 
We present the formula of the isotope
coefficient on the basis of the spin-fluctuation theory, and
show that the isotope effect observed by experiments is consistent
with this formula.

\section{Hamiltonian}

The total Hamiltonian is the sum of the electronic part, the phonon part and
the electron-phonon interaction part:
\begin{equation}
H = H_{el}+H_{ph}+H_{el-ph}.
\end{equation}
Each term in the Hamiltonian is given as follows.

We adopt that the ferromagnetism arises from the on-site Coulomb
interaction and use the Hubbard model
given as
\begin{equation}
H_{el}= \sum_{{\bf k}\sigma}\xi_{{\bf k}}c_{{\bf k}\sigma}^{\dag}
c_{{\bf k}\sigma}+U\sum_i n_{i\uparrow}n_{i\downarrow}, 
\end{equation}
where $c_{{\bf k}\sigma}$ and $c_{{\bf k}\sigma}^{\dag}$ are Fourier
transforms of the annihilation and creation operators $c_{i\sigma}$ 
and $c_{i\sigma}^{\dag}$
at the site $i$, respectively.
$n_{i\sigma}=c^{\dag}_{i\sigma}c_{i\sigma}$ is the number operator, and
$U$ is the strength of the on-site Coulomb interaction.
$\xi_{{\bf k}}=\epsilon_{{\bf k}}-\mu$ is the dispersion relation
measured from the chemical potential $\mu$.
The phonon part of the Hamiltonian is given by
\begin{equation}
H_{ph}= \sum_{{\bf k}}\omega_{{\bf k}}\left(b_{{\bf k}}^{\dag}b_{{\bf k}}
+\frac{1}{2}\right),
\end{equation}
where $b_{{\bf k}}$ and $b_{{\bf k}}^{\dag}$ are operators for the phonon
and $\omega_{{\bf k}}$ is the phonon dispersion.
The electron-phonon interaction is\cite{agd}
\begin{equation}
H_{el-ph}= \gamma \int d^3x\sum_{\sigma}\psi_{\sigma}^{\dag}({\bf x})
\psi_{\sigma}({\bf x})\varphi({\bf x}),
\end{equation}
where the electron field $\psi_{\sigma}$ and the phonon field $\varphi$
are defined, respectively, as follows:
\begin{eqnarray}
\psi_{\sigma}({\bf x})&=& \frac{1}{\sqrt{V}}\sum_{{\bf k}}e^{i{\bf k}\cdot
{\bf x}}c_{{\bf k}\sigma},\\
\varphi({\bf x}) &=& \frac{1}{\sqrt{V}}\sum_{{\bf k}}\left(
\frac{\hbar\omega_k}{2}\right)^{1/2}\left( b_{{\bf k}}e^{i{\bf k}\cdot{{\bf x}}}
+b_{{\bf k}}^{\dag}e^{-i{{\bf k}}\cdot{{\bf x}}}\right),
\end{eqnarray}
where $V$ is the volume of the system.

\section{Electron-phonon vertex correction}

\begin{figure}[htbp]
\begin{center}
\includegraphics[height=3.1cm]{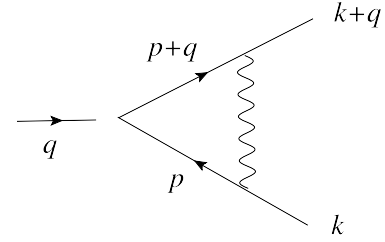}
\caption{
Electron-Phonon vertex function.  
The wavy line indicates the phonon propagator.
The momenta $k$, $p$ and $q$ represent 4-momenta such as
$k=(i\epsilon_m,{\bf q})$,
$p=(i\omega_n,{\bf p})$ and $q=(i\nu_{\ell},{\bf q})$.
}
\end{center}
\end{figure}

\subsection{Electron-phonon vertex function}

The electron-phonon vertex correction plays an important role in the
isotope effect in itinerant ferromagnets. 
The vertex function $\Gamma(k,q+k;q)$, shown in Fig.1,
is written as
\begin{eqnarray}
\Gamma(k,k+q;q)&=& -\gamma^2\frac{1}{\beta}\sum_n\int\frac{d^3p}{(2\pi)^3}
G_0(i\omega_n,{\bf p})\nonumber\\
&\times& G_0(i\omega_n+i\nu_{\ell},{\bf p}+{\bf q})
D_0(i\omega_n-i\epsilon_m,{\bf p}-{\bf k}),\nonumber\\
\end{eqnarray}
where $G_0$ is the electron Green function and $D_0$ is the phonon Green
function\cite{agd}:
\begin{eqnarray}
G_0(i\omega_n,{\bf p})&=& \frac{1}{i\omega_n-\xi_{\bf p}},\\
D_0(i\nu_{\ell},{\bf k})&=& \frac{\omega_{{\bf k}}^2}{(i\nu_{\ell})^2
-\omega_{{\bf k}}^2},
\end{eqnarray}
where $\omega_{{\bf k}}$ is the phonon dispersion relation.
It is known as the Migdal theorem that the vertex correction is of order
of $\omega_D/\epsilon_F$\cite{agd,mig58,fett,her76,fay79}.
The vertex function is evaluated by using the method of Green function 
theory\cite{agd,fay79,eli63}.
In the limit $(i\nu,{\bf q})\rightarrow 0$, we obtain\cite{fay79}
\begin{equation}
\Gamma(k,k+q;q)\simeq -\gamma^2\rho(0)\frac{1}{2}
\frac{\omega_{\rm D}}{\epsilon_{\rm F}}
\ln\left(\frac{\epsilon_{\rm F}}{\omega_{\rm D}}\right).
\label{vertexap}
\end{equation}

\begin{figure}[htbp]
\begin{center}
\includegraphics[width=6cm]{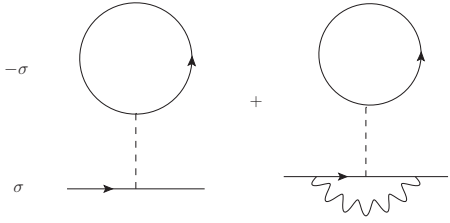}
\caption{
Lowest-order electron self-energy corrections.  
The second term is that due to the vertex correction.
The dashed line indicates the Coulomb interaction $U$ and the wavy line
shows the phonon propagator. 
}
\end{center}
\end{figure}

We consider the self-energy corrections shown in Fig.2.
The first term is the Hartree term that stems from the on-site Coulomb
interaction and the second one includes the vertex correction.
When we use the approximation in eq.(\ref{vertexap}), the self-energy
is
\begin{equation}
\Sigma_{{\bf k}\sigma}= Un_{-\sigma}
+Ug\rho(0)\frac{\omega_{\rm D}}{2\epsilon_{\rm F}}
\ln\left(\frac{\epsilon_{\rm F}}{\omega_{\rm D}}\right) n_{-\sigma},
\end{equation}
where $g=-\gamma^2$ and 
the number density of electrons with spin $\sigma$ is denoted as
$n_{\sigma}=(1/N)\sum_i\langle n_{i\sigma}\rangle$ where $N$ is the
number of sites.

\subsection{Electron susceptibility}

We show contributions to the electron susceptibility $\chi^{+-}$
in Fig. 3.  They are given by  
\begin{equation}
\chi^{(a)}(0)= \sum_k G_0(k)G_0(k)
\Gamma(k,k;q=0),
\end{equation}
\begin{eqnarray}
\chi^{(b)}(0)&=& \sum_k \sum_q
G_0(k)G_0(k+q)G_0(k)G_0(k+q)\nonumber\\
&&\times U \Gamma(k,k+q;q),
\end{eqnarray}
where $\Sigma_k$ indicates $\Sigma_k=(1/\beta)\sum_m\int d^dk/(2\pi)^d$.
The term in Fig. 3(b) contains the electron-phonon
vertex correction as well as the Coulomb interaction.
We show the vertex function $\Gamma(k,k+q;q)$ for small $q\simeq 0$ as a
function of $\omega_D/t$ in Fig.4.  We put $k_0=0$ and 
${\bf k}=(\pi,0,0)$ and ${\bf k}=(\pi/2,\pi/2,\pi/2)$ in three
dimensions and ${\bf k}=(\pi,0)$ in two dimensions.
The vertex function is of the order of $\omega_D/t$ when $\omega_D/t$
is small, $\omega_D/t\ll 1$, in accordance with the Migdal
theorem\cite{agd,fett}.
The result shows the same behavior regardless of space dimension in two- 
and three-dimensional cases.
 
We evaluated $\chi^{(a)}$ and $\chi^{(b)}$ in Fig.3 in two dimensions.
%since the vertex function is almost independent of dimension.
We show them in Fig.5 as a function of $\omega_D/t$ where the upper line
indicates $\chi^{(a)}/g$ and the lower one is for $\chi^{(b)}/Ug$.
The result indicates that
$\chi^{(b)}/Ug$ is smaller than $\chi^{(a)}/g$ by about two orders of magnitude.

\subsection{Two-particle interaction}

It was pointed out that the vertex correction to susceptibility
$\chi^{+-}$
in Fig.3(b) may give a large contribution to the isotope effect
when we include the effective electron-hole two-particle interaction shown in 
Fig.6\cite{app80}.
We examine this here.
The two-particle interaction in Fig.6 is denoted as $\Pi(p_1,p_2;p_3.p_4)$.
The susceptibility in Fig.3(b) with the two-particle interaction
is written as (Fig.7)
\begin{eqnarray}
\chi^{(b)}_{e-p}(Q)&=& \sum_k \sum_p \sum_q
G(k)G(k+q)G(k+Q)G(k+q+Q)
\nonumber\\
&\times& (-\gamma^2)G(p)G(p+q)D(k-p)\nonumber\\
&\times& \Pi(p,k+Q;p+q,k+q+Q),
\end{eqnarray}
where $k=(i\epsilon_m,{\bf k})$, $q=(i\nu_{\ell},{\bf q})$,
$p=(i\omega_n,{\bf p})$ and $Q=(iQ_0,{\bf Q})$.
$G$ and $D$ are Green's functions including the interaction corrections.
For the on-site Coulomb interaction, the effective electron-hole interaction
reads
\begin{equation}
\Pi(p_1,p_2;p_3,p_4)= \frac{U}{1-U\chi(p_1-p_2)},
\end{equation}
where $\chi(q)$ is the electron susceptibility.
We consider the case $Q=0$.
For the ferromagnetic case, $\chi^{(b)}_{e-p}(Q)$ is less than the value
obtained by approximating the two-particle interaction $\Pi$ by $U/(1-U\chi(0))$
since $\chi$ may have a peak at $p_1-p_2=0$. 
%\begin{eqnarray}
%\chi^{(b)}_{e-p}(0) &\leq& \frac{1}{\beta}\sum_m\frac{d^dk}{(2\pi)^d}
%\frac{1}{\beta}\sum_{\ell}\frac{d^dq}{(2\pi)^d} \nonumber\\
%&\times& G(k)G(k+q)G(k)G(k+q)\Gamma(k,k+q;q)\nonumber\\
%&\times& U/\left( 1-U\chi(0) \right).
%\end{eqnarray}
The contribution in Fig. 3(b) is enhanced by the factor $U/(1-U\chi(0))$.
When $U$ is near the critical value $U_c$, for example, $1-U\chi_0\sim -0.1$, 
the term from $\chi^{(b)}$ is still small compared to that from $\chi^{(a)}$.
When $U$ is extremely near $U_c$ such as $1-U\chi_0\sim -0.01$, 
the problem becomes delicate.
We do not, however, consider this region in this paper because a more
precise theory is needed to investigate the critical region.

\begin{figure}[htbp]
\begin{center}
\includegraphics[height=3cm]{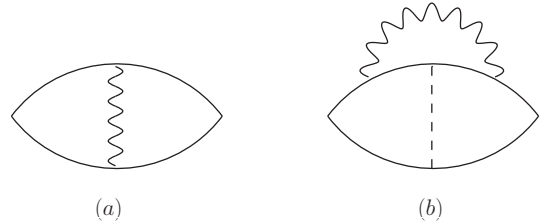}
\caption{
Contributions to the susceptibility $\chi^{+-}$ with the electron-phonon 
vertex correction. 
They are denoted as $\chi^{(a)}$ and $\chi^{(b)}$.
}
\end{center}
\label{chi-ph}
\end{figure}

\begin{figure}[htbp]
\begin{center}
\includegraphics[height=6cm]{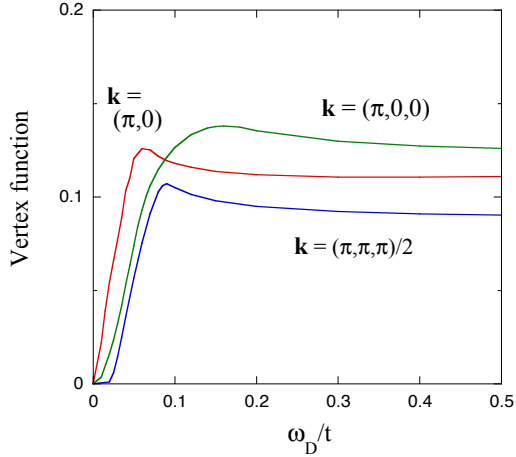}
\caption{
Vertex function $\Gamma(k,k+q;q)$ as a function of $\omega_D/t$ when 
$q=(iq_0,{\bf q})$ is small in the static
limit $k_0=q_0=0$. 
The wave number ${\bf k}$ is ${\bf k}=(\pi,0,0)$ and
${\bf k}=(\pi/2,\pi/2,\pi/2)$ in three dimensions ($\L\times L\times L$
lattice) and ${\bf k}=(\pi,0)$
in two dimensions ($L\times L$ lattice).
The electron dispersion is $\xi_{{\bf k}}= -2t(\cos(k_x)+\cos(k_y)+\cos(k_z))-\mu$
and $\xi_{{\bf k}}=-2t(\cos(k_x)+\cos(k_y))-\mu$.
We put $L=200$ and the summation with respect to the Matsubara frequency
is restricted to $-n_{\beta}\le n\le n_{\beta}$ with $n_{\beta}=100$ 
and the inverse temperature $\beta=8/t$.
In actual calculations ${\bf q}$ is kept finite such as
${\bf q}=(0.01\pi,0.01\pi,0.01\pi)$.
}
\end{center}
\end{figure}

\begin{figure}[htbp]
\begin{center}
\includegraphics[height=6cm]{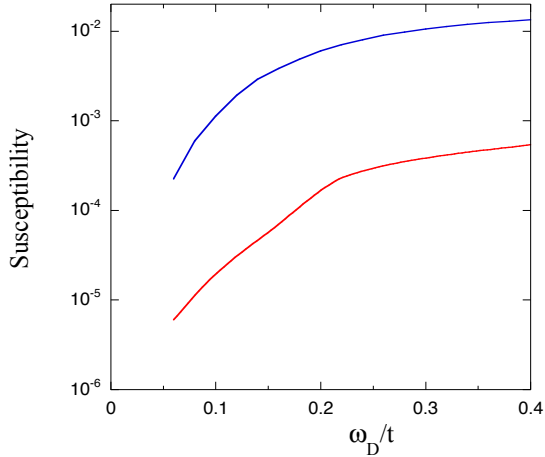}
\caption{
Susceptibilities $\chi^{(a)}/g$ in Fig.3(a) (upper) and $\chi^{(b)}/Ug$
in Fig.3(b) (lower), respectively, 
as a function of $\omega_D/t$. 
}
\end{center}
\end{figure}

\begin{figure}[htbp]
\begin{center}
\includegraphics[width=9.0cm]{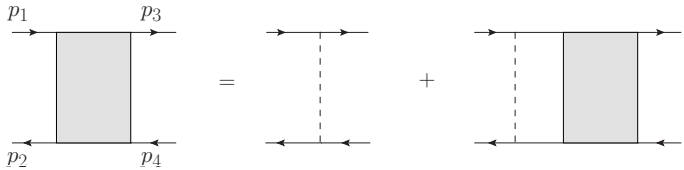}
\caption{
Bethe-Salpeter equation for the effective two-particle
electron-hole interaction 
$\Pi(p_1,p_2;p_3,p_4)$ with $p_1-p_2=p_3-p_4$\cite{fett}.
}
\end{center}
\label{ladder-u}
\end{figure}

\begin{figure}[htbp]
\begin{center}
\includegraphics[width=6.0cm]{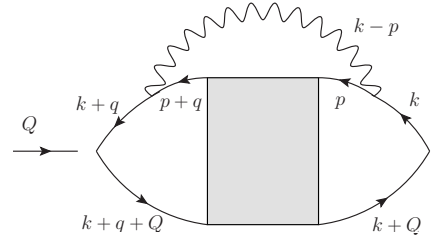}
\caption{
Susceptibility with the effective two-particle interaction and the
electron-phonon vertex correction. 
}
\end{center}
\label{chi-ld-vx}
\end{figure}

\section{Isotope effect in itinerant ferromagnets}

\subsection{Isotope effect in the Stoner theory\cite{app80}}

The magnetization
$\Delta\equiv n_{\uparrow}-n_{\downarrow}$ is determined form the
mean-field equation given by
\begin{equation}
n_{\uparrow}-n_{\downarrow}=\frac{1}{V}\sum_{{\bf k}}
(f(E_{{\bf k}\uparrow})-f(E_{{\bf k}\downarrow})),
\end{equation}
where $E_{{\bf k}\sigma}=\xi_{{\bf k}}+\Sigma_{{\bf k}\sigma}$
and $f(E)$ is the Fermi distribution function.
The equation is written as
up to the order of $\Delta$:
\begin{equation}
\Delta= -U_{{\rm eff}}\Delta\int d\xi\rho(\xi)f'(\xi),
\end{equation}
where
\begin{equation}
U_{{\rm eff}}= U+Ug\rho(0)\frac{\omega_D}{2\epsilon_F}\ln
\left(\frac{\epsilon_F}{\omega_D}\right).
\end{equation}
The Curie temperature 
$T_{\rm c}$ in the mean-field theory is\cite{mor85}
\begin{equation}
k_{\rm B}T_{\rm c}= \sqrt{A}\sqrt{1-\frac{1}{U_{{\rm eff}}\rho(0)}},
\end{equation}
where $A$ is a constant.  
This result is also obtained from the condition
$1/2=U\chi(0)|_{T=T_c}$ in the RPA theory.
Because the contribution in Fig. 3(b) is small compared with that
in Fig. 3(a) (at least except the region just near the critical value of $U$),
we neglect the term in Fig. 3(b).
Because $\omega_{\rm D}\propto 1/\sqrt{M}$ where
$M$ is an corresponding atomic mass, we obtain
\begin{equation}
\frac{\partial\ln T_{\rm c}}{\partial\ln M}= -\frac{1}{4}
\frac{1}{U_{{\rm eff}}\rho(0)-1}\frac{\partial\ln U_{{\rm eff}}}
{\partial\ln\omega_{\rm D}}.
\end{equation}
From this relation, we obtain positive derivative 
$\partial T_{\rm c}/\partial M>0$.
The isotope coefficient $\alpha=-\partial\ln T_c/\partial\ln M$ 
is\cite{app80}
\begin{equation}
\alpha= -\frac{1}{4}\frac{U}{U_{\rm eff}(U_{\rm eff}\rho(0)-1)}|g\rho(0)|
\frac{\omega_D}{2\epsilon_F}\Big[ 
\ln\left(\frac{\epsilon_F}{\omega_D}\right)-1 \Big].
\end{equation}

Let us estimate the shift of $T_{\rm c}$ by using this formula.
For $T_{\rm c}=160$K, $M=16$ and $\Delta M=2$, 
$\Delta T_{\rm c}= -T_{\rm c}\Delta M/M\cdot\alpha$ is
\begin{equation}
\Delta T_{\rm c}\simeq 0.25\frac{1}{U_{\rm eff}\rho(0)-1}|g\rho(0)|,
\end{equation}
where the unit is K (kelvin).
We obtain $\Delta T_{\rm c}\simeq 0.075$K for $U_{\rm eff}\rho(0)=2$
and $\Delta T_{\rm c}\simeq 0.15$ for $U_{\rm eff}\rho(0)=1.5$ where 
we set $|g\rho(0)|=0.3$.
$U$ should be very close to the critical value of $U$, like
$U\rho(0)\sim 1.07$, to agree
with the observation $\Delta T_{\rm c}\simeq 1$K.

\subsection{Self-consistent spin-fluctuation theory}

The physical properties of weak itinerant ferromagnets are well
understood by the self-consistent renormalization (SCR) theory of spin
fluctuation\cite{mor85}.
We must take account of spin fluctuation to evaluate the isotope shift
of $T_c$.
We use the SCR theory for this purpose.
Let us consider the free-energy functional of an $S^4$ theory: 
\begin{eqnarray}
F&=& \sum_{\bf q}\left( \frac{1}{2\chi_0({\bf q})}-U\right)
|{\bf S}_{\bf q}|^2
+\frac{\lambda}{4}\sum_j|{\bf S}_j|^4-{\bf S}_{{\bf q}=0}\cdot{\bf h}
\nonumber\\
&=& \sum_{\bf q}\left( \frac{1}{2\chi_0({\bf q})}-U\right)
|{\bf S}_{\bf q}|^2
+ \frac{\lambda}{4}\frac{1}{N}\sum_{qq'q''}({\bf S}_{\bf q}\cdot
{\bf S}_{-{\bf q}'})\nonumber\\
&\times& ({\bf S}_{{\bf q}''}\cdot{\bf S}_{{\bf q}'-{\bf q}''-{\bf q}})
 -{\bf S}_{{\bf q}=0}\cdot{\bf h},
\end{eqnarray}
where ${\bf S}_j$ is the spin density and 
${\bf h}$ is the magnetic field.
$N$ indicates the number of lattice sites.
The Fourier decomposition of ${\bf S}_j$ is defined by 
\begin{equation}
{\bf S}_j= \frac{1}{\sqrt{N}}\sum_{\bf q}{\bf S}_{\bf q} 
e^{i{\bf q}\cdot{\bf R}_j}.
\end{equation}
$\lambda$ is the coupling constant that indicates the strength of the
mode-mode couplings, and $\chi_0({\bf q})$ is the susceptibility of the
non-interacting system.
We adopt that ${\bf S}_{{\bf q}=0}=(0,0,S)$ and ${\bf h}=(0,0,h)$ where $S=S(T)$ 
is the magnetization (order parameter).
From the equation $\partial F/\partial S=0$,
the susceptibility $\chi(T)\equiv S/h$ is given by
\begin{equation}
\frac{1}{\chi(T)}= \frac{1}{\chi_0}-2U+\lambda (3m_{\parallel}^2+2m_{\perp}^2)
+\lambda S^2.
\end{equation}
$\chi_0=\chi_0({\bf q}=0)$ is the uniform susceptibility and is
assumed to be temperature independent: $\chi_0=\rho(0)/2$.
From the fluctuation-dissipation theorem, $m_{\alpha}^2$ is given as
\begin{eqnarray}
m_{\alpha}^2&=& \frac{2}{\pi}
\int \frac{d^3q}{(2\pi)^3} \int_0^{\infty}d\omega 
\left( \frac{1}{2}+\frac{1}{e^{\omega/T}-1}\right)
{\rm Im}\chi_{\alpha}({\bf q},\omega)\nonumber\\
&=& (m_{\alpha}^2)_{\rm zp}+(m_{\alpha}^2)_{\rm th}.
\end{eqnarray}
At $T>T_{\rm c}$ we have $S=0$ so that we can assume $m_x^2=m_y^2=m_z^2$.
At $T=T_{\rm c}$ we obtain
\begin{equation}
\frac{1}{\lambda}\left( 2U-\frac{1}{\chi_0}\right)= 
5\left(m_{\alpha}^2\right)\Big|_{T=T_c}.
\end{equation}

We include the electron-phonon correction in the susceptibility $\chi({\bf q},\omega)$:
\begin{equation}
\chi({\bf q},\omega)= \chi_{el}({\bf q},\omega)+\chi_{e-ph}({\bf q},\omega),
\end{equation}
where $\chi_{el}$ is the susceptibility without the electron-phonon correction and 
$\chi_{e-ph}$ is of order $\omega_D/\epsilon_F$ coming from the diagram in Fig. 3(a).
When we use an approximation in eq.(10), $\chi$ is approximated as
\begin{equation}
\chi({\bf q},\omega)\simeq \chi_{el}({\bf q},\omega)
\left( 1+g\rho(0)\frac{\omega_D}{2\epsilon_F}\ln\frac{\epsilon_F}{\omega_D}\right).
\label{chiep}
\end{equation}
We use the following form for the susceptibility 
$\chi_{el}({\bf q},\omega)$\cite{mor85,hert76,mil93}:
\begin{equation}
\frac{1}{\chi_{el}({\bf q},\omega)}= \frac{1}{\chi_{el}(0,0)}+Aq^2-iC\frac{\omega}{q},
\end{equation}
where $A$ and $C$ are constants, and $\chi_{el}(0,0)=\chi_{el}(T)$.
The electron-phonon interaction gives a correction of order of 
$\omega_D/\epsilon_F$.
Then at $T=T_{\rm c}$, $(m_{\alpha}^2)_{\rm th}$ is proportional to 
$T_{\rm c}^{4/3}(1+c_0 g\rho(0)\omega_D/(2\epsilon_F)\ln(\epsilon_F/\omega_D))$ 
with a constant $c_0$.
In the approximation in eq.(\ref{chiep}), we have $c_0=1$.
Numerical calculations in Fig. 5 indicate that $c_0$ is small, especially
for small $\omega_D/t$,
due to multiple integrals of momenta.

We substitute $U_{\rm eff}$ to $U$ to take account of the electron-phonon 
interaction. 
The zero-point fluctuation $(m_{\alpha}^2)_{\rm zp}$ is simply a constant
at $T=T_{\rm c}$ and we include this contribution in $U_{\rm eff}$.
This results in a formula for the isotope 
coefficient $\alpha=-\partial\ln T_{\rm c}/\partial\ln M$ given as 
\begin{equation}
\alpha= -\frac{3}{8}
\left( \frac{U\rho(0)}{U_{{\rm eff}}\rho(0)-1}-c_0\right) |g\rho(0)|
\frac{\omega_{\rm D}}{2\epsilon_{\rm F}}
\biggl[ \ln\left(\frac{\epsilon_{\rm F}}{\omega_{\rm D}}\right)-1\biggr].
\end{equation}
$\partial \ln T_c/\partial\ln M$ is positive when 
$U_{\rm eff}\rho(0)/(U_{\rm eff}\rho(0)-1)-c_0 >0$.
This inequality holds as far as $c_0<1$ for $U_{\rm eff}\rho(0)>1$.

%\begin{figure}[htbp]
%\begin{center}
%\includegraphics[width=7cm]{dtc-u}
%\caption{
%The Curie temperature shift $\Delta T_{\rm c}$ as a function of $U\rho$ where
%the solid curve is for the self-consistent spin-fluctuation theory and
%the dotted curve is for the Stoner theory.
%We set $\delta_{\rm zp}=0$.
%The lower dashed-dotted curve indicates $U_{{\rm eff}}\rho-1$.
%We set $T_{\rm c}\simeq 160$K, $M=16$, $\Delta M=2$, $D/\epsilon_{\rm F}=2$,
%$\epsilon_{\rm F}/\omega_{\rm D}=25$ and $|g\rho|=0.2$.
%The vanishing of $U_{{\rm eff}}\rho-1$ indicates that the ferromagnetic transition
%occurs.
%}
%\end{center}
%\end{figure}

For SrRuO$_3$, the Debye frequency is $\omega_D\simeq 340$cm$^{-1}\sim 490$K.
Then we set $\omega_D/\epsilon_F\sim 0.05$.
For $T_{\rm c}=160{\rm K}$, $M=16$ and $\Delta M=2$, we obtain
$\Delta T_{\rm c}=-T_{\rm c}\Delta M/M\cdot \alpha$ as
\begin{equation}
\Delta T_{\rm c} \simeq 0.374 \left( \frac{U\rho(0)}{U_{\rm eff}\rho(0)-1}
-c_0\right)|g\rho(0)|,
\end{equation}
in units of K.
This formula gives the value which agrees with experimental results.
For example,
we have $\Delta T_{\rm c}\simeq 0.22$K for $U_{\rm eff}\rho(0)=2$
and $\Delta T_{\rm c}\simeq 0.34$K for $U_{\rm eff}\rho(0)=1.5$, where
$g\rho(0)=-0.3$ and we neglect $c_0$.
$\Delta T_{\rm c}$ increases as $U$
approaches the critical value.
The experimental value $\Delta T_{\rm c}\simeq 1$K is obtained
when $U\rho(0)\simeq 1.13$.

\section{Summary}

We have presented a theory of the isotope effect of Curie temperature $T_c$
in itinerant ferromagnets.
It is primarily important to determine the sign of the shift of $T_{\rm c}$
for isotope substitution.
Our picture is that the decrease of the Debye frequency results in the
increase of relative strength of the Coulomb interaction and this 
leads to a positive shift of $T_{\rm c}$ as $M$ increases.

The isotope shift of $T_{\rm c}$ occurs through the electron-phonon coupling.
This effect is of order of $\omega_{\rm D}/\epsilon_{\rm F}$ because the 
electron-phonon
interaction is restricted to the region within an energy shell of thickness 
$\omega_{\rm D}$.
We have presented the formula on the basis of the spin-fluctuation theory.
The isotope shift of $T_{\rm c}$ is obtained as a function of the
electron-phonon coupling $g$ and the on-site Coulomb interaction $U$.
These are not determined within a theory, and are treated as parameters.
The sign of the isotope shift $\Delta  T_{\rm c}$ of the Curie 
temperature agrees with the experimental result and $\Delta T_{\rm c}$
decreases as $U$ increases.
The experimental value of $\Delta T_{\rm c}$ is consistent with the
formula if
we adopt that $U$ is not far from the critical value of the ferromagnetic
transition. This assumption is reasonable for usual itinerant 
ferromagnetic materials.

Our theory cannot be applied to materials such La$_{1-x}$Ca$_x$MnO$_3$
because the ferromagnetic transition is caused by the double-exchange 
interaction and the Jahn-Teller effect may play a role in these
ferromagnets.

Numerical calculations were performed at the Supercomputer Center of the
Institute for Solid State Physics, University of Tokyo.

\end{document}